\documentclass[conference]{IEEEtran}
\IEEEoverridecommandlockouts
\usepackage{cite}
\usepackage{amsmath,amssymb,amsfonts}
\usepackage{graphicx}
\usepackage{textcomp}
\usepackage{xcolor}
\def\BibTeX{{\rm B\kern-.05em{\sc i\kern-.025em b}\kern-.08em
    T\kern-.1667em\lower.7ex\hbox{E}\kern-.125emX}}
\usepackage{graphicx} 
\usepackage{amsmath,amsfonts}
\usepackage{algorithm}
\usepackage{algpseudocode}
\usepackage{array}
\usepackage[caption=false,font=normalsize,labelfont=sf,textfont=sf]{subfig}
\usepackage{textcomp}
\usepackage{stfloats}
\usepackage{url}
\usepackage{verbatim}
\usepackage{graphicx}
\usepackage{cite}
\usepackage{dsfont}
\usepackage{xcolor}
\DeclareMathOperator{\rank}{rank}
\usepackage{subfig}
\usepackage{acronym} \input{Acronym}
\begin{document}

\title{
RIS-Assisted Rank Enhancement With Commodity WiFi Transceivers: Real-World Experiments
\thanks{This work was supported by the German Research Foundation (“Deutsche Forschungsgemeinschaft”) (DFG) under Project–ID 449601577.}
}
\author{
\IEEEauthorblockN{Aymen Khaleel and Aydin Sezgin
}
\IEEEauthorblockA{
\textit{Ruhr University Bochum, Bochum, Germany}  \\
Email: \{aymen.khaleel, aydin.sezgin\}@rub.de
}
}

\maketitle

\begin{abstract}
Reconfigurable intelligent surfaces (RISs) are a promising enabling technology for the sixth-generation ($6$G) of wireless communications. RISs, thanks to their intelligent design, can reshape the wireless channel to provide favorable propagation conditions for information transfer. In this work, we experimentally investigate the potential of RISs to enhance the effective rank of \ac{MIMO} channels, thereby improving spatial multiplexing capabilities. In our experiment, commodity WiFi transceivers are used, representing a practical MIMO system. In this context, we propose a passive beam-focusing technique to manipulate the propagation channel between each transmit-receive antenna pair and achieve a favorable propagation condition for rank improvement. The proposed algorithm is tested in two different channel scenarios: low and intermediate ranks. Experimental results show that, when the channel is rank-deficient, the RIS can significantly increase the rank by $112\%$ from its default value without the RIS, providing a rank increment of $1.5$. When the rank has an intermediate value, a maximum of $61\%$ enhancement can be achieved, corresponding to a rank increment of $1$. These results provide the first experimental evidence of RIS-driven rank manipulation with off-the-shelf WiFi hardware, offering practical insights into RIS deployment for spatial multiplexing gains.

\end{abstract}

\begin{IEEEkeywords}
RIS, MIMO, rank, spatial multiplexing, channel
\end{IEEEkeywords}
\acresetall
\section{Introduction}
{R}{econfigurable} intelligent surfaces (RISs) are an emerging technology with a high potential to be integrated into almost all of the existing wireless communications systems. Due to its unique features such as passiveness, intelligent reconfiguration, and relatively low implementation cost \cite{RIS_Advantages}, RISs are one of the main enabling technologies envisioned for the sixth-generation (6G) of wireless communications \cite{RISfor6G}.


Comprised of a large number of electronically controllable elements, RISs promise to reshape the transmission medium to create favorable propagation conditions for wireless information transfer \cite{dataset}. In this context, the authors in \cite{chanl-meas} conducted real-world field trials and reported time-domain channel measurements of a RIS-assisted \ac{SISO} system. The authors showed that the RIS can reduce the delay spread by focusing the energy of the reflection paths, mitigating the multipath effect. In \cite{mimo-chnl-limit}, the authors considered a \ac{MIMO} system assisted by beyond-diagonal RIS, where they investigated the limits to which the RIS can reengineer the channel. In particular, the authors investigated the impact of RIS on communication degrees of freedom (DoF), singular value spread, power gain, and capacity.

In wireless networks, \ac{MIMO} systems are the dominating implemented technology for high data rate and multiple user access through spatial multiplexing. To achieve this goal, a \ac{MIMO} channel needs to have enough DoF. Here, the effective rank appears as a reliable measurement to quantify the number of independent spatial dimensions the channel can effectively utilize, which is defined by the effective number of non-negligible singular values associated with the channel matrix \cite{eff-rank}. The authors in \cite{emil-rank} showed theoretically that the RIS can be used to transfer a rank-deficient channel into a full-rank channel by properly adjusting the RIS to create additional distinctive propagation paths. In \cite{rank-tera}, the authors considered the use of the RIS to enhance the spatial properties of ultra-massive \ac{MIMO} systems for terahertz communications. By optimizing the antenna and RIS elements placement, the authors reported an enhancement on the order of hundreds in the channel rank. In \cite{rank-cope}, the authors considered a multi-user \ac{MIMO} cooperative communication system, where the RIS is used to overcome the channel rank-deficiency and manage inter-user interference. In \cite{rank-drone}, considering sub-$6$ GHz, $28$ GHz, and $60$ GHz frequency bands, the authors showed that, by optimizing their placement, RISs can effectively enhance the rank for the drone-assisted air-to-ground channel. In \cite{rank-dist}, the authors considered a multi-user \ac{MISO} system where they showed that using distributed RISs achieves superior performance in enhancing the channel rank compared to the single RIS case. In \cite{Dof}, considering both active and passive RISs, the authors investigated the DoF region problem for the time-selective $K$-user interference channel. In \cite{spatial}, the authors considered an uplink multi-user system assisted by an RIS, where they studied the spatial multiplexing performance for a linear receiver.

Experimentally, few studies have investigated the impact of RIS on the channel's spatial characteristics. Specifically, in \cite{rank-urb-exp}, the authors experimentally tested the RIS effect to maximize the channel gain at the \ac{UE} side for a sub-$6$ GHz \ac{MIMO} channel in an urban macrocell, achieving a capacity enhancement of $50\%$. The authors also reported measurements of the channel spatial characteristics impacted by the RIS. In \cite{rank-exp1}, the authors showed in a lab experiment, using universal software radio peripheral (USRP) devices, that the RIS can be utilized to enhance the effective rank in a $2\times 2$ \ac{MIMO} system, where an enhancement of $30\%$ is reported.

Against this background, this work is the first to experimentally investigate the impact of reconfigurable intelligent surfaces (RISs) on channel rank using commodity WiFi transceivers. In contrast to prior works that used software-defined radios such as USRPs, this experiment provides practical engineering insights into RIS deployment in real-world wireless networks and characterizes their interactions with off-the-shelf commercial devices. In this context, we propose a passive beam focusing technique at the RIS side to manipulate the propagation channel to enhance the rank. Furthermore, this work distinguishes itself from the state-of-the-art in \cite{rank-exp1} from the following aspects. It considers a larger setup of $3\times 3$ \ac{MIMO} system at the $5$ GHz band with a large RIS size of $1024$ elements. The investigation of RIS impact is more comprehensive, considering the two scenarios: low and intermediate rank. Finally, a copper sheet was used as a reference to illustrate what the RIS can achieve beyond simply reflecting incoming signals, providing practical engineering insight into its performance limits.


The rest of the paper is organized as follows: Section II introduces the system model, followed by the proposed solution algorithm in Section III. Section IV provides the experimental setups, followed by the obtained measurements and their discussions. Finally, in Section V, the paper's main conclusions are given.

\textit{Notation:} Matrices and column vectors are denoted by an upper and lower case boldface letters, respectively. $\mathbf{X} \in \mathbb{C}^{m \times k}$ denotes a complex-valued matrix $\mathbf{X}$ with $m \times k$ size, where $\mathbf{X}^T$ is the transpose and $\lfloor\mathbf{X}\rfloor_{n,h}$ is the $(n,h)$-th entry. $\mathbf{0}_N$ and $\mathbf{I}_N$ are the $N$-dimensional all-zeros column vector, and the $N \times N$ identity matrix, respectively.
$x \sim \mathcal{CN}(0, q^2)$ stands for complex Gaussian distributed random variable (RV) with mean $\mathbb{E}[x] = 0$ and variance $\text{Var}[x] = q^2$. Big-O notation $\mathcal{O}(\cdot)$ is used for representing the computational complexity of algorithms.
\begin{figure}[t!]
    \centering
    \includegraphics[width=70mm]{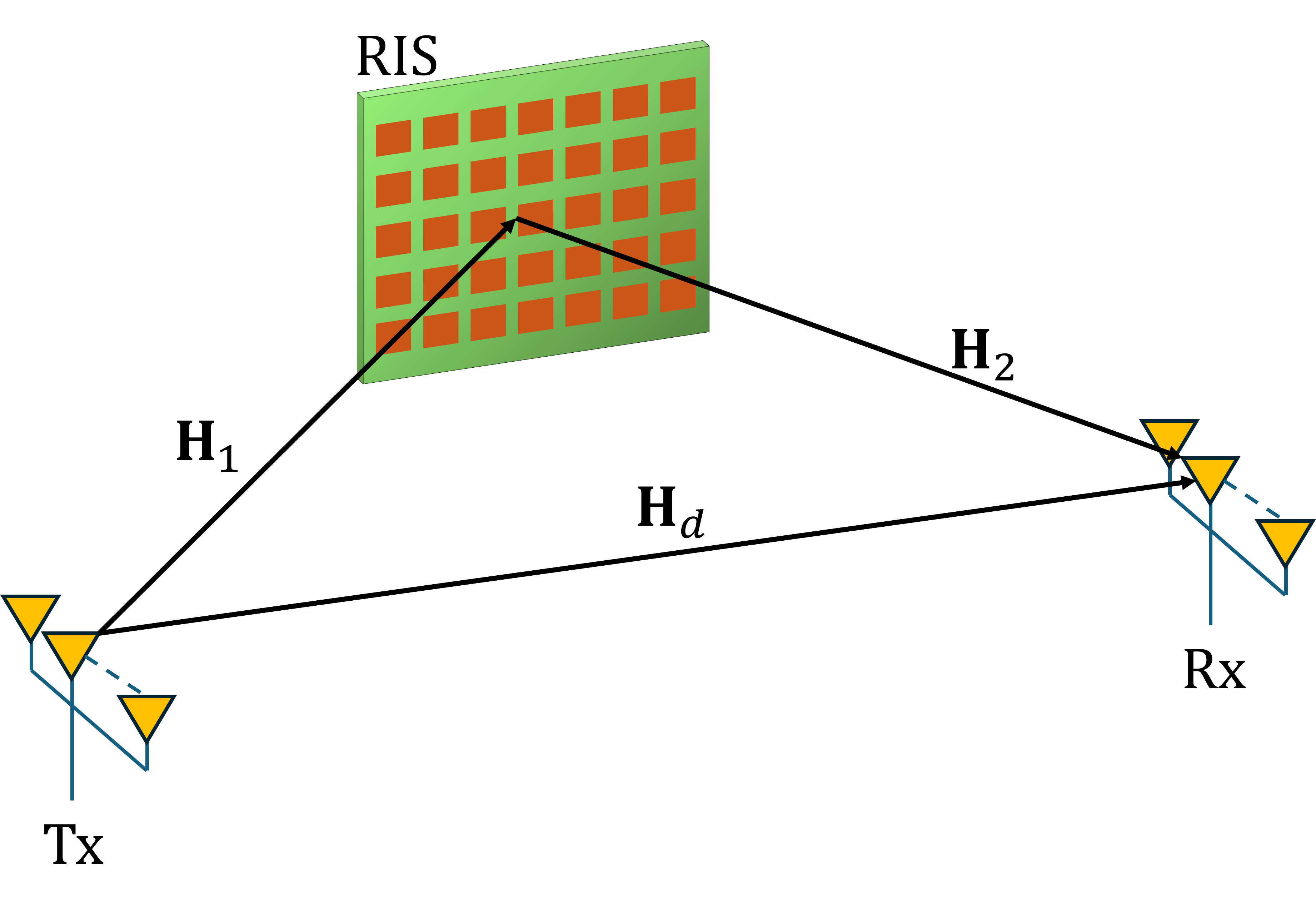}
    \caption{A RIS-assisted MIMO system.}\label{fig:system_block}
\end{figure}
\section{System Model}
Consider a $N_T\times N_R$ MIMO system, where $N_T$ and $N_R$ denote the number of transmit and receive antennas, respectively, assisted by a RIS of $N$ elements, see Fig. \ref{fig:system_block}. Accordingly, the received signal vector $\mathbf{y}\in\mathbb{C}^{N_R\times 1}$ can be given as
\begin{equation}
    \mathbf{y} = \mathbf{H} \mathbf{V} \mathbf{x} + \mathbf{n}, \label{eq:1}
\end{equation}
where $\mathbf{x}\in\mathbb{C}^{N_T\times 1}$ is the transmitted signal vector and $\mathbf{V}\in\mathbb{C}^{N_T\times N_T}$ is the precoding matrix. $\mathbf{H}=\mathbf{H}_2^T \mathbf{\Psi} \mathbf{H}_1 + \mathbf{H}_d$, with  $\mathbf{H}_d\in\mathbb{C}^{N_R\times N_T}$ is the transmitter (Tx)-receiver (Rx) channel matrix; $\mathbf{H}_1\in\mathbb{C}^{N\times N_T}$ is the Tx-RIS channel matrix, $\mathbf{H}_1=[\mathbf{h}_{1,1},\dots,\mathbf{h}_{1,n_T}\dots, \mathbf{h}_{1,N_T}]$, where $\mathbf{h}_{1,n_T}\mathbb{C}^{N\times 1}$ is the ($n_T$-th Tx antenna)-RIS channel vector. Similarly, $\mathbf{H}_2\in\mathbb{C}^{N\times N_R}$ is the RIS-Rx channel matrix, $\mathbf{H}_2=[\mathbf{h}_{2,1},\dots,\mathbf{h}_{2,n_R}\dots, \mathbf{h}_{2, N_R}]$, where $\mathbf{h}_{2,n_R}\mathbb{C}^{N\times 1}$ is the RIS-($n_R$-th) Rx antenna channel vector. $\mathbf{n}\in\mathbb{C}^{N_R \times 1}$ is the additive white Gaussian noise (AWGN) samples vector, $\mathbf{n}\sim\mathcal{CN}(\mathbf{0}_{N_R},\sigma^2\mathbf{I}_{N_R})$. The RIS is represented by the diagonal phase shift matrix $\mathbf{\Psi}\in\mathbb{C}^{N\times N}$, where $\lfloor\mathbf{\Psi}\rfloor_{n,n}=\beta_ne^{j\psi_n}, \forall n\in{1,\dots,N}$, with $\beta_n$ and $\psi_n$ are the reflection amplitude and phase applied by the $n$-th element, respectively.

Considering the singular value decomposition (SVD) of $\mathbf{H}=\mathbf{U}\mathbf{\Sigma}\mathbf{V}$, with the singular values $q_i\geq q_{i+1}, i=1, \dots, \rank(\mathbf{H})-1$, then the channel effective rank is given by \cite{eff-rank}
\begin{align}
    R_e=\text{exp}\left(\sum_{i=1}^{\rank(\mathbf{H})}-\bar{q}_i \ln{\bar{q_i}}\right),\label{eq:rank}
\end{align}
where $\bar{q_i}=q_i/\sum_i q_i$. In the following section, we introduce our proposed approach to maximize $R_e$.

Note that, due to the RIS size and operating frequency considered in this experiment, both the Tx and Rx are located within the near-field region of the RIS. This will be further explained in Section IV.
\section{RIS-Side Passive Beam Focusing}

It is worth noting that the proposed algorithm in this section is implemented on the RIS side without any hardware modification to the Tx-Rx system. Likewise, on the software level, only limited feedback control is required from the Rx to choose the proper phase configuration at the RIS side. Nevertheless, no transmission/recaption protocol/signaling modification to the current WiFi standard is made. This gives a clear practical advantage for the proposed algorithm to be readily integrated into the legacy WiFi system.

Due to near-field propagation and the associated spherical wavefront curvature, the channel responses of different $n_T$--$n_R$ antenna pairs are geometrically coupled through the RIS aperture and share a structured, element-dependent Fresnel phase pattern across its surface \cite{near-field-ris}.  In this way, optimizing the RIS phase shifts for a specific $n_T-n_R$ antenna pair also partially aligns nearby pairs, coherently boosting multiple entries of the MIMO channel matrix. As a result, not only the largest, but several singular values increase together, improving the matrix conditioning and resulting in a smaller condition number \cite{emil-rank}. Accordingly, in Algorithm 1, we propose a low-complexity passive beam-focusing strategy at the RIS side, searching over a structured set of RIS configurations to identify the one that maximizes $R_e$ in \eqref{eq:rank}.
\begin{algorithm}
\caption{Passive Beam Focusing}
\begin{algorithmic}[1]
\State \textbf{Input:}  $N_T,N_R, N$.
\State \textbf{Output:}  $\mathbf{\Psi}$.
\State Define $G=|\mathbf{h}_{2,n_R}^T\mathbf{\Psi}\mathbf{h}_{1,n_T}|^2$.
\For{$n_T = 1$ to $N_T$}
    \For{$n_R = 1$ to $N_R$}
\State $\psi_n=0, \forall n\in\{1, \dots, N\}$.
\State Compute $G_0=G$.
        \For{$n = 1$ to $N$}
          \State $\psi_n=\pi$
          \State Compute $G$.
              \If{$G<G_0$}
                \State $\psi_n=0$.
                \Else
                \State $G_0=G$.
             \EndIf
        \State Compute the SVD of $\mathbf{H}$ and obtain \\$\;\quad\quad\;\;$ $R_e(n_T,n_R)$ using \eqref{eq:rank}. 
        \EndFor
    \EndFor
\EndFor
\State Obtain $(n_T^*,n_R^*)=\underset{n_T,n_R}{\arg\max}\quad R_e(n_T,n_R)$.
\State  \textbf{Return} $\psi_n^*,\forall n$, associated with the antenna pair \\$\;\quad\quad\quad\quad\;$ ($n_T^*,n_R^*$).
\end{algorithmic}
\end{algorithm}
Specifically, in each iteration of Algorithm 1, a specific $n_T-n_R$ antenna pair is considered, where the phase shifts of the RIS elements are adjusted to constructively/destructively align the reflected signal paths between these two antennas. To achieve this goal, alternating optimization is considered \cite{alter} (Steps $8$-$15$) to obtain the required element state ($0$ or $\pi$) for the discrete phase shift RIS used in this experiment, enabling constructive alignment. Note that, to align the reflected signals destructively, the inequality sign (Step 11) needs to be reversed. Next, at the Rx side, the SVD of the overall channel $\mathbf{H}$, which is now reshaped by the obtained $\mathbf{\Psi}$, is obtained to compute the effective rank using \eqref{eq:rank}. Spanning all of the $n_T-n_R$ possible combinations, the antenna pair $n_T^*-n_R^*$ associated with the maximum $R_e$ is feed-backed to the RIS control circuit. Finally, the RIS applies the $\psi^*_n$ associated with $n_T^*-n_R^*$ to achieve the maximum effective rank, $R_e^*$. In practice, Algorithm 1 can be frequently applied whenever the channel condition changes in order to reshape the signal propagation again to guarantee the maximum achievable effective rank. In this context, \ac{CSI}  needs to be available at the RIS side to compute $G$. Nevertheless, considering the \ac{LoS} links between the transmitter/receiver and the RIS, only the angles of arrival/departure need to be acquired, effectively limiting the required feedback to the RIS over several channel coherence intervals.

It can be seen that the computational complexity of Algorithm $1$ is dominated by the nested search over all $(n_T,n_R)$ antenna pairs, resulting in $N_TN_R$ loop iterations. Each iteration incurs a linear computational cost of $\mathcal{O}(N)$. In addition, within each iteration a SVD computation of $\mathcal{O}\!\left(\min(N_T,N_R)^2\max(N_T,N_R)\right)$ operations is required for each antenna pair. Consequently, the overall computational complexity is given as
\begin{align}
\mathcal{O}\!\left(N_TN_RN + N_TN_R\min(N_T,N_R)^2\max(N_T,N_R)\right).
\end{align}

\section{Experimental Setup and Measurements}
In this section, we first introduce the experimental setup, and then the reported measurements are discussed.
\subsection{Experimental Setup}
\begin{figure}[t!]
\centering
\subfloat[]{
    \includegraphics[width=.237\textwidth]{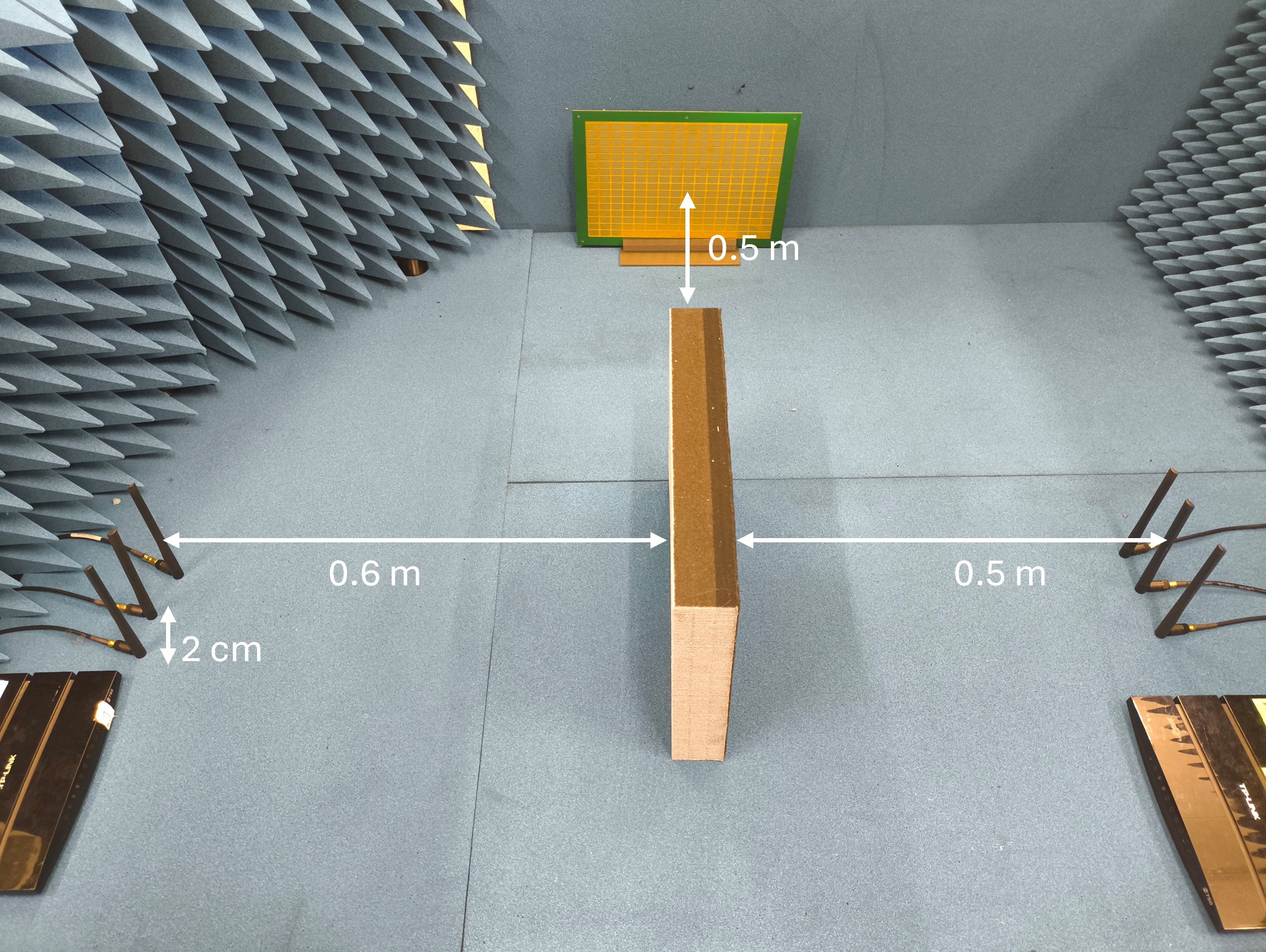}}
\hfill\subfloat[]{
    \centering
    \includegraphics[angle=90,width=.225\textwidth]{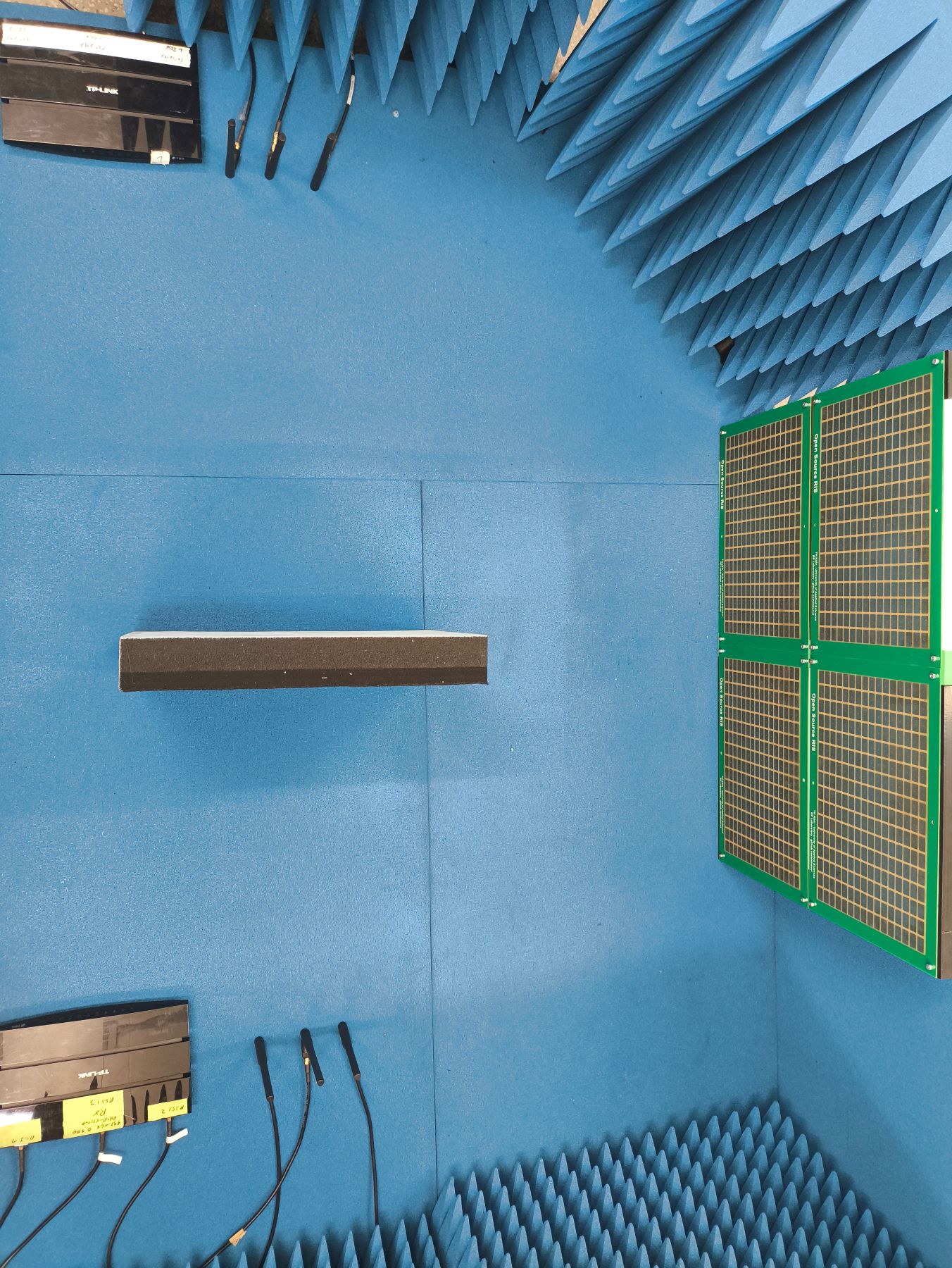}
   } \caption{RIS-assisted MIMO setup with: (a) $1$ RIS and (b) $4$ RISs.}\label{fig:setup-ris}
\end{figure}

Two Atheros ATH9k-based TP-Link N750 commodity WiFi transceivers are used as an $N_T\times N_R$ MIMO system that uses the IEEE 802.11n standard, with $N_T=N_R=3$. To align with the operating frequency of the RIS, WiFi channel $48$ is used, with a $20$ MHz bandwidth and a center frequency of $5.24$ GHz. Note that the Tx and Rx are located within the near-field region of the RIS: for $\lambda=0.0573$ m and $N=256, 1024$, we get the Rayleigh distance $N\lambda/2=6.5$ m and $27.5$ m, respectively, which is much larger than the Tx-RIS and RIS-Rx distances shown in Fig. \ref{fig:setup-ris}(a). To enable \ac{CSI} extraction, the two routers are patched using the Atheros CSI tool \cite{csi_tool} and run the open-source operating system OpenWrt \cite{openwrt}. Here, the transmitter operates in injection mode, where neither router is connected to an access point or a client network. Specifically, the transmitter router injects packets over the wireless channel, while the receiver router receives these packets and extracts the CSI from them. A host computer is used to access OpenWrt on both routers and to forward the received CSI from the receiver to MATLAB for processing. In this context, the channel matrix is first constructed for a single subcarrier, and then SVD is used to obtain the singular values. Next, using \eqref{eq:rank}, the effective rank is calculated and averaged over $100$ channel realizations.

\begin{figure}[t!]
    \subfloat[]{
    \includegraphics[width=41mm,height=33mm]{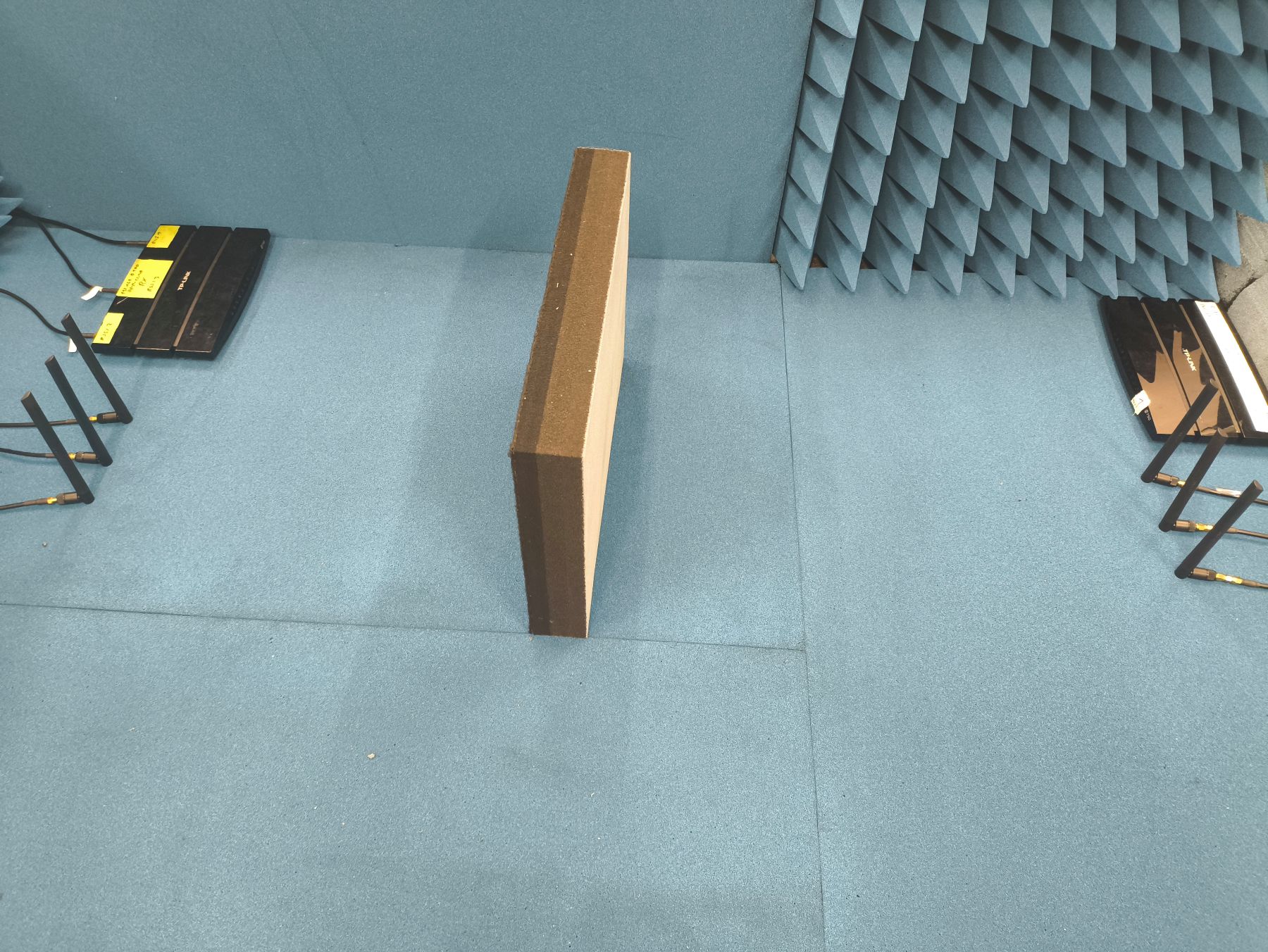}}
 \label{fig:setup}
\hfill
\subfloat[]{    \includegraphics[width=44.5mm,height=33mm]{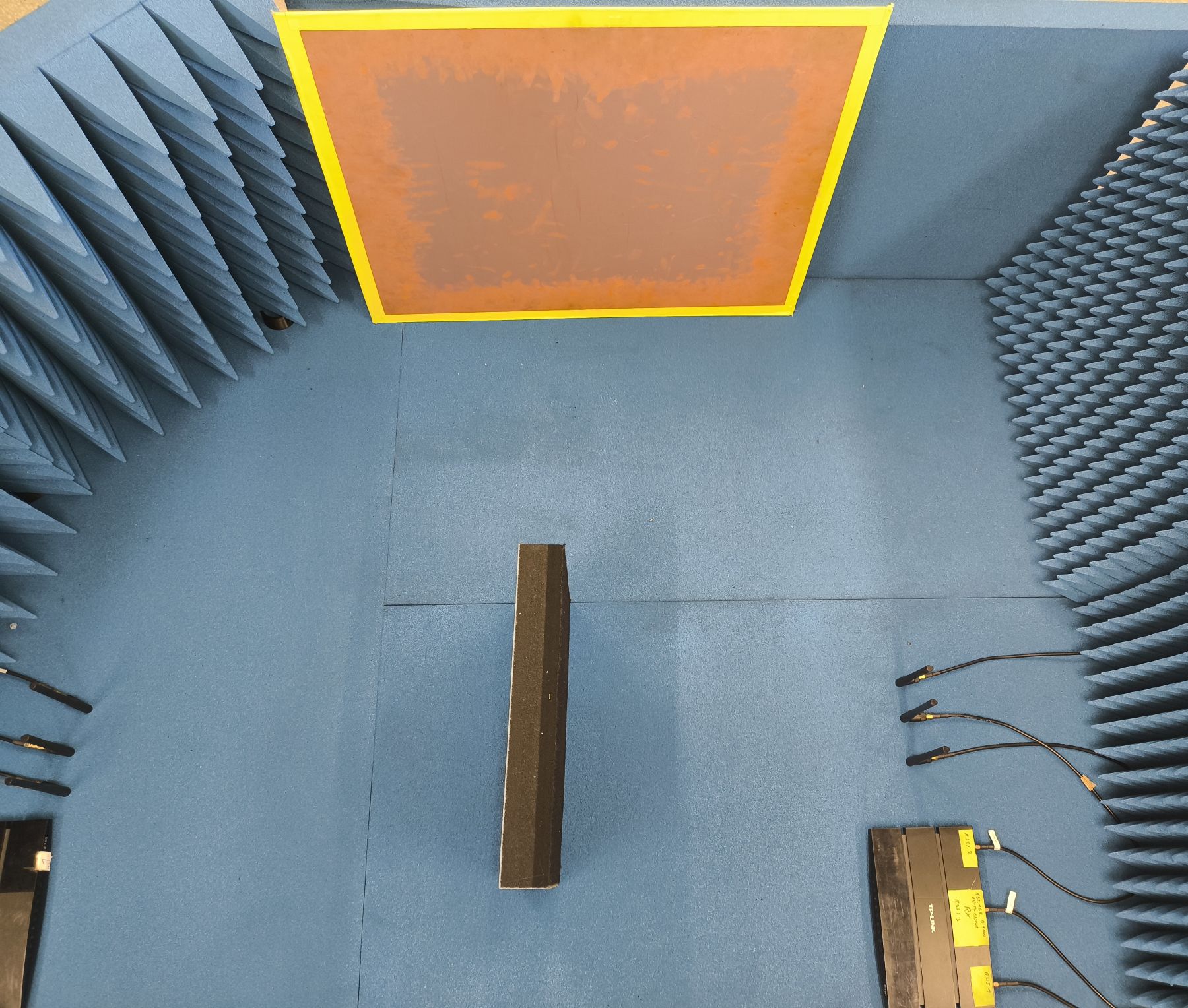}}
    \caption{A $3\times 3$ \ac{MIMO} system without RISs: (a) default setup and (b) with a copper sheet. }\label{fig:setup-def}
\end{figure}

Fig. \ref{fig:setup-def}(a) shows the default setup without RISs, where the isolation foam is used to control the propagation of the signal to obtain the required rank at the default state. Here, by modifying the propagation environment through changing the isolation foam position between the transmitter and receiver, we obtain the channel rank values that serve as the baseline for our comparison when the RISs are used later. Furthermore, we use ``Copper Sheet" with a size that is $9$ times the area of the used RIS, as a reference to the performance of the RIS in enhancing the channel rank, see Fig. \ref{fig:setup-def}(b).
\begin{table*}[h!]
\caption{Low $R_e$}
\centering
\begin{tabular}{|l|c|c|c|c|}
\hline
\textbf{Method} & $R_e$  ($1$ RIS) & \textbf{Difference \%} & $R_e$  ($4$ RISs) &\textbf{Difference \%}  \\
\hline
w/o RIS & 1.2745 & -& 1.2745 & - \\
Passive Beam Focusing     & 1.9394 & 52.1680 & 2.8082 & 112.8082\\
Fixed Phase           & 1.9651 & 54.1827 & 1.7337 & 31.3800 \\
Copper Sheet              & 1.9171 & 50.4156  & 1.9171 & 50.4156\\
\hline
\end{tabular}\label{table:low}
\end{table*}
\begin{table*}[h!]
\caption{Intermediate $R_e$}
\centering
\begin{tabular}{|l|c|c|c|c|}
\hline
\textbf{Method} & $R_e$  ($1$ RIS) & \textbf{Difference \%} & $R_e$  ($4$ RIS) &\textbf{Difference \%} \\
\hline
w/o RIS & 1.7713 & - &1.7713 & - \\
Passive Beam Focusing       & 2.5896 & 46.1977 & 2.8661 & 61.8066 \\
Fixed Phase           & 2.5903 & 46.2393& 2.5456 & 43.7110 \\
Copper Sheet              & 2.4746 & 39.7036& 2.4746 & 39.7036 \\
\hline
\end{tabular}\label{table:medium}
\end{table*}
To manipulate the propagation channel, RIS modules with $1$-bit phase shift control are used to provide two phase shift levels, $0$ and $\pi$, with a total of $N=256$ elements for each module. The used RIS hardware module is fully detailed in \cite{ris-module}. 
Two different scenarios are considered to illustrate the MIMO channel behavior affected by the RIS, with low and intermediate ranks, respectively. Specifically, to realize these rank values, we modify the experiment setup by changing the antennas' geometrical deployment and using isolation foam. 

Next, we add the RIS to the setup to examine its effect in all two scenarios, where we use $1$ and $4$ RIS modules to illustrate the impact of RIS size, as shown in Fig. \ref{fig:setup-ris}. Here, we consider different RIS phase shift configurations, as follows. In ``Fixed Phase", all the elements are adjusted either to $0$ or $\pi$. In ``Passive Beam Focusing", the RIS phase shifts are adjusted according to Algorithm 1 to provide beam focusing at the receiver side by, respectively, enabling constructive and destructive alignment of the reflected signals.

\subsection{Experimental Measurements}
In Table \ref{table:low}, we show the scenario when the channel $R_e$ is low and the RIS is used to increase it. Using a single RIS, the 
three methods, ``Fixed Phase", ``Passive Beam Focusing", and ``Copper Sheet", achieve similar performance, which can be explained as follows. Here, we compute $\text{Difference}=(R_e-\bar{R}_e)/\bar{R}_e$, where $\bar{R}_e$ is the effective rank computed at the default state without the RISs/copper sheet. First, it can be seen that adjusting the phase shifts of individual elements to achieve passive beam focusing has a negligible impact on the rank value compared to just fixing the phase for all of them. Second, despite the large size of the copper sheet, using the RIS achieves a slightly better performance, which can be explained as follows. The flat copper sheet enforces specular reflection, resulting in an approximately separable Tx-RIS-Rx response, where the energy is concentrated into a single dominant mode. This affects the largest singular value while leaving secondary singular values small, resulting in poor channel conditioning and lower rank gain that is highly sensitive to geometry. In contrast, as a discrete finite aperture, the RIS provides a Fresnel-like field with multiple partially coherent lobes across the elements array. This boosts several matrix entries, improves the minimum singular value, and delivers a more robust rank. Using $4$ RISs, a significant enhancement can be noticed using Algorithm 1, where their effective rank has increased by approximately $1.5$, providing $112\%$ increment to the default value without the RIS. In contrast, fixing the phases for all elements achieved less performance compared to the single RIS case, which can be explained as follows. As the size increases, the RIS array factor narrows and the response becomes more specular/separable. This makes signal
energy concentrate into a single dominant spatial mode, which amplifies the largest singular value while secondary ones shrink relative to it, resulting in a higher conditioning number and less rank. Specifically, the larger size makes the edge-diffraction ``randomness", which makes the propagation more scatterer-rich, averaged out. This shows that using the RIS intelligently (beam focusing) reshapes the propagation channel more effectively beyond the reflecting surface area with fixed (not optimized) phases. 

In Table \ref{table:medium}, we consider the scenario where the channel effective rank has an intermediate value of $1.7$. Overall, except the ``Fixed Phase" with $4$ RISs, all of the considered methods lead to a lower rank improvement compared to the low rank scenario discussed before, Table \ref{table:low}. This can be explained by the propagation structure of the channel, which is now much scatter-richer, inherently limiting the impact of the considered methods to provide a dramatic change, as was the case previously, Table \ref{table:low}. Nevertheless, for the $4$ RISs case, the beam focusing method has superior performance, increasing the rank by approximately $1$, showing a $61.8\%$ increment from the default value without the RIS. Furthermore, fixing the phase for all elements performs less compared to the beam focusing method, showing that the optimization of the elements' phases has a critical effect on the channel rank. The same behavior appears when comparing the previously mentioned methods with the ``Copper Sheet". 

Overall, it can be seen that Algorithm 1 has superior performance compared to the one in \cite{rank-exp1}, where a maximum enhancement of $30\%$ is reported in that work.

\section{Conclusion}
In this work, commodity WiFi transceivers are considered, representing a $3\times 3$ MIMO system, to investigate the use of RISs to impact the propagation channel by measuring the effective rank. Two different methods are considered for adjusting the phases of the RIS: the fixed phase method and the beam focusing method. As a baseline, a copper sheet is used to assess the performance of the RIS relative to it. Experimental measurements demonstrate that the RIS can significantly reshape the propagation channel due to its discrete surface structure, and this effect is more pronounced with intelligent phase adjustment. Specifically, when the propagation channel is not rich-scattered (low rank), the RIS can dramatically increase the rank through passive beam focusing. The RIS impact decreases when the channel propagation paths are moderately scattered (intermediate rank), leaving fewer degrees of freedom for the RIS to exploit.

\bibliographystyle{IEEEtran}
\bibliography{refs}

\end{document}